# Valence, exchange interaction, and location of Mn in polycrystalline $Mn_xGa_{1-x}N$ (x≤0.04)

A. Furrer[1], K. W. Krämer[2], A. Podlesnyak[3], V. Pomjakushin[1], D. Sheptyakov[1], and O. V. Safonova[4]

[1] Laboratory for Neutron Scattering, Paul Scherrer Institut, CH-5232 Villigen PSI, Switzerland

[2] Department of Chemistry and Biochemistry, University of Bern, CH-3012 Bern, Switzerland

[3] Neutron Scattering Division, Oak Ridge National Laboratory, Oak Ridge, Tennessee 37831, USA

[4] Paul Scherrer Institut, CH-5232 Villigen PSI, Switzerland

We present an experimental study for polycrystalline samples of the diluted magnetic semiconductor $Mn_xGa_{1-x}N$ (x≤0.04) in order to address some of the existing controversial issues. Different techniques were used to characterize the electronic, magnetic, and structural properties of the samples, and inelastic neutron scattering was employed to determine the magnetic excitations associated with Mn monomers and dimers. Our main conclusions are as follows: (i) The valence of the Mn ions is 2+. (ii) The $Mn^{2+}$ ions experience a substantial single-ion axial anisotropy with parameter D=0.027(3) meV. (iii) Nearest-neighbor $Mn^{2+}$ ions are coupled antiferromagnetically. The exchange parameter J=-0.140(7) meV is independent of the Mn content x, *i.e.*, there is no evidence for hole-induced modifications of J towards a potentially high Curie temperature postulated in the literature.

There is an ongoing debate on the physical properties of the compound $Mn_xGa_{1-x}N$, a diluted magnetic semiconductor (DMS) with potential applications in spintronics and blue-LED technologies [1,2]. The interest in $Mn_xGa_{1-x}N$ is driven by the prediction of Mn-induced ferromagnetism with Curie temperatures $T_C$ exceeding room temperature [3], which is required for a technological breakthrough in the field of DMSs. Up to the present, myriads of experiments have been performed for $Mn_xGa_{1-x}N$ compounds, but the conclusions still remain highly controversial as recently summarized by Nelson *et al*. [4]. A basic problem with $Mn_xGa_{1-x}N$ is the low solubility of Mn ions in the host compound GaN, so that the investigated samples are often contaminated by Mn clusters or other phases which are ferromagnetic in nature, *e.g.*, MnGa ($T_C$>600 K) and $Mn_4N$ (ferrimagnetic, $T_C$=738 K). For this reason, the observation of ferromagnetism above room temperature reported in the literature has to be considered with caution, especially as none of these findings has resulted in a device working at room temperature. There are some other important questions associated with $Mn_xGa_{1-x}N$ for which so far no common agreement has been obtained, namely: (i) What is the valence of the Mn ions ($Mn^{2+}$ *vs* $Mn^{3+}$)? (ii) What is the nature and the size of the magnetic exchange interaction (ferromagnetic *vs* antiferromagnetic)? (iii) Where are the Mn ions located (regular Ga positions *vs* interstitial positions)?

It is the purpose of the present work to provide answers to all these questions through different experiments on polycrystalline samples of $Mn_xGa_{1-x}N$. The first part of this Letter describes the detailed characterization of the samples, which is an indispensable issue, since some of the existing controversies are due to the lack of information on the sample properties. We continue by presenting inelastic neutron scattering (INS) investigations of the magnetic excitations associated with Mn monomers and dimers for which so far no information is available in the literature. As a result of the different experimental techniques applied to $Mn_xGa_{1-x}N$, we arrive at consistent conclusions as summarized in the abstract.

Based on the procedure outlined by Szyszko *et al.* [5], polycrystalline $Mn_xGa_{1-x}N$ samples were synthesized with manganese concentrations up to x=0.08. The Mn content x was determined by energy dispersive X-ray (EDX) spectroscopy. Analyses by X-ray and neutron diffraction showed that $Mn_xGa_{1-x}N$ crystallizes in the hexagonal space group *P6₃mc*, but even for low values of x an impurity phase of type $MnN_y$ (y<1) with tetragonal space group *I4/mmm* was always present. We concentrated our experimental study on detailed investigations of two samples with $x_1$=0.024(3) and $x_2$=0.072(2) from EDX. The fractional weight of the impurity phase determined by X-ray diffraction amounted to 0.4(3)% and 3.7(5)%, respectively. Similar values 0.38(7)% and 2.95(15)%, respectively, were obtained from neutron diffraction. Thus, the actual Mn content x of the main phase $Mn_xGa_{1-x}N$ has to be corrected accordingly, *i.e.*, for $x_1$=0.024(3) and $x_2$=0.072(2) we have $Mn_{0.02}Ga_{0.98}N$ and $Mn_{0.04}Ga_{0.96}N$, with typically 10% uncertainty for x.

The neutron powder diffraction experiments were performed with use of the high-resolution diffractometer for thermal neutrons HRPT (λ=1.155 Å, high resolution mode with δd/d=$10^{-3}$) [6] at the spallation neutron source SINQ at PSI Villigen. The refinements of the crystal structures were carried out with the program FULLPROF [7]. The diffraction data taken at T=293 K were analyzed both for the main phase $Mn_xGa_{1-x}N$ and for the impurity phase $MnN_y$. In model I the dopant Mn ions were treated as substitutional ions at the Ga-site with position (2/3,1/3,0). However, the refinement provided an occupation number of Mn close to zero. Consequently, in further data refinements we allowed the z-coordinate of the Mn position (1/3,2/3,z) to be varied, resulting in a z-displacement of about 0.3 Å from the regular Ga site for model II. Model III was triggered by the results of a Fourier analysis, based on the difference between the observed and the calculated structure factors, which gave evidence for additional scattering from Mn at the interstitial site (2/3,1/3,z) with z=0.18(3). The structure parameters obtained for the three models are summarized in Table I.

Models I and III are based on fixed z-coordinates of the Mn ions. If we start the refinement with the initial parameters of models I and III by releasing the z-constraint of the Mn position, the fitting procedure converges to the parameters of model II. We therefore feel that model II provides the most probable solution for the structure of $Mn_xGa_{1-x}N$. Another argument for model II (and against model III) is the fact that in principle there is no room for placing ions at interstitial sites in the densely packed Wurtzite crystal structure of $Mn_xGa_{1-x}N$, unless either N or Ga vacancies are created in order to comply with the interatomic distances. This is in contrast to the related DMS $Mn_xGa_{1-x}As$, where the dopant Mn ions can occupy interstitial sites commensurate with the zinc-blende crystal structure [8]. Due to the small differences of the reliability factors listed in Table I for the three models, there remains some uncertainty concerning the z-coordinate of the Mn ions. However, it is important to realize that the local symmetry of Mn ions being either at the sites (2/3,1/3,z) or at regular Ga sites (2/3,1/3,0) is identical with tetrahedral coordination, which is an important aspect for the analysis of both the X-ray absorption near-edge structure (XANES) and the INS experiments described below.

XANES experiments at the Mn K-edge (6.539 keV) were performed at room temperature with use of the instrument SuperXAS [9] at the Swiss Light Source (SLS) at PSI Villigen. A 2.9 T super-bend magnet provided X-rays, and a Si collimating mirror at 2.5 mrad rejected the higher harmonics. A channel-cut Si(111) monochromator was used to select the desired photon energy. A Rh-coated toroidal mirror focused the beam to 0.5 and 0.1 mm in horizontal and vertical dimensions, respectively. For the calibration of the beam energy we used an Fe foil (Fe K-edge at 7.112 keV). $Mn_2O_3$ and MnO powders served as reference samples for $Mn^{3+}$ and $Mn^{2+}$ in octahedral coordination, respectively. All the samples were measured in the transmission mode. Figure 1 shows the normalized Mn K-edge XANES data recorded for the four samples, which can be interpreted as follows: (i) For $Mn_xGa_{1-x}N$ with x=0.02 the energy of the K-edge is very close to that of MnO, which favors $Mn^{2+}$. (ii) Raising the Mn content to x=0.04 slightly increases the oxidation state of the Mn ions (probably due to the increased weight of the impurity phase $MnN_y$), but it still remains close to 2+. (iii) The strong pre-edge features around 6.54 keV suggest that the Mn ions have tetrahedral coordination [10], in contrast to the reference samples $Mn_2O_3$ and MnO with octahedral Mn coordination. Similar results were obtained from XANES experiments performed for $Mn_xGa_{1-x}N$ (0.03<x<0.09) layers grown by molecular beam epitaxy on [0001] SiC substrates [11].

Electron paramagnetic resonance (EPR) experiments performed with a microwave frequency of 9.39 GHz gave rise to a resonance at a magnetic field of 0.33 T, similar to the results described by Zajac *et al.* [12], which agrees with the field-induced splitting of the $m=|\pm 1/2\rangle$ ground state of $Mn^{2+}$ ions for g=2. In addition, analyses of magnetic susceptibility and magnetization data also favor $Mn^{2+}$.

INS experiments were carried out with use of the high-resolution time-of-flight spectrometer CNCS [13] at the spallation neutron source (SNS) at Oak Ridge National Laboratory. The samples were enclosed in aluminum cylinders of 8 mm diameter and placed into a He cryostat to achieve temperatures T≥1.7 K. Additional experiments were performed for vanadium to allow the correction of the raw data with respect to background, detector efficiency, and absorption according to standard procedures. We searched for magnetic excitations over a wide energy range, but we did not observe magnetic intensity for energy transfers >0.6 meV. Energy spectra taken for $Mn_xGa_{1-x}N$ with x=0.04 in the neutron energy-gain configuration are shown in Fig. 2(a) for moduli of the scattering vector **Q** in the range 0.5≤Q≤1.5 Å$^{-1}$. The energy of the incoming neutrons was 1.55 meV, with instrumental energy resolutions of Gaussian shape increasing from 20 µeV to 29 µeV for energy transfers from 0 meV to 0.5 meV, respectively. We observe an increase of the intensity upon raising the temperature from 1.7 K to 6.0 K. This becomes more clear by plotting the difference of the energy spectra as shown in Fig. 2(b), which has the advantage that uncertainties about the background are automatically eliminated. Taking intensity differences has proven to be an extremely powerful procedure to analyze INS data [14].

The data of Fig. 2(b) exhibit three partially resolved lines, which were analyzed by Gaussians without any constraints in the least-squares fitting procedure, except for fixing the background at zero intensity. The results are shown as full and broken curves in Fig. 2(b). We interpret the three lines in terms of Mn multimer transitions associated with $Mn_xGa_{1-x}N$. We can neglect the scattering contributions from the impurity phase $MnN_y$. MnN and $Mn_3N_2$ order antiferromagnetically below very high Néel temperatures $T_N$=660 K and $T_N$=920 K, respectively, with nearest-neighbor exchange parameters of the order of -20 meV [15], giving rise to a spin-wave density-of-states far above the energy window covered by the present INS experiments.

For the low Mn content only monomers and dimers have to be considered, which based on a random distribution of x Mn ions over the positions (1/3,2/3,z) occur with probabilities $p_M=(1-x)^{12}$ and $p_D=6x(1-x)^{18}$, respectively. The linewidths are considerably enhanced beyond the instrumental energy resolution due to local structural effects [16]. A proper identification of the lines is possible by considering both the spin Hamiltonian and the neutron cross-section for monomers and dimers.

The spin Hamiltonian of Mn monomers is given by

$$H = D(s^z)^2 , \qquad (1)$$

where D is the axial single-ion anisotropy parameter, and $s^z$ denotes the z-component of the spin operator **s** of the Mn ions. The corresponding neutron cross-section for monomer transitions $|m\rangle \rightarrow |m'\rangle$ is defined by [17]

$$\frac{d^2\sigma}{d\Omega d\omega} \propto F^2(Q) n_m(T) \left| T_1^{\Delta m} \right|^2 , \qquad (2)$$

where F(Q) is the magnetic form factor, $n_m(T)$ the Boltzmann population factor of the initial state $|m\rangle$, and $-s^z \leq m \leq s^z$. The transition matrix element $\left| T_1^{\Delta m} \right|$ gives rise to the selection rules

$$\Delta m = m - m' = 0, \pm 1 . \qquad (3)$$

The spin Hamiltonian of Mn dimers is given by

$$H = -2 J_1 \mathbf{s_1} \cdot \mathbf{s_2} + D \left[ \left( s_1^z \right)^2 + \left( s_2^z \right)^2 \right] , \qquad (4)$$

where J is the bilinear exchange parameter. It is convenient to base the diagonalization of Eq. (4) on the dimer states $|S,M\rangle$, where $\mathbf{S}=\mathbf{s_1}+\mathbf{s_2}$ is the total spin and $-S \leq M \leq S$. For $Mn^{2+}$ ($Mn^{3+}$) ions with $s_i=5/2$ ($s_i=2$), ferromagnetic (J>0) and antiferromagnetic (J<0) exchange give rise to an S=5 (S=4) and an S=0 (S=0) ground state, respectively. The anisotropy term has the effect of splitting the spin states $|S\rangle$ into the substates $|S,\pm M\rangle$. The low-energy level schemes for $Mn^{2+}$ and $Mn^{3+}$ monomers and antiferromagnetically coupled dimers are illustrated in Fig. 3.

The neutron cross-section for dimer transitions $|S,M\rangle \rightarrow |S',M'\rangle$ is defined by [18]

$$\left.\frac{d^2\sigma}{d\Omega d\omega}\right|_{\Delta M=0} \propto F^2(Q) n_{|S,M\rangle}(T) \left\{ \frac{2}{3} + (-1)^{\Delta S} \left[ \frac{2\sin(QR)}{(QR)^3} - \frac{2\cos(QR)}{(QR)^2} \right] \right\} \left| T_1^{\Delta M=0} \right|^2$$

(5)

$$\left.\frac{d^2\sigma}{d\Omega d\omega}\right|_{\Delta M=\pm 1} \propto F^2(Q) n_{|S,M\rangle}(T) \left\{ \frac{2}{3} - (-1)^{\Delta S} \left[ \frac{2\sin(QR)}{(QR)^3} - \frac{2\cos(QR)}{(QR)^2} - \frac{\sin(QR)}{QR} \right] \right\} \left| T_1^{\Delta M=\pm 1} \right|^2$$

where R is the distance between the two dimer spins. The transition matrix element $|T_1^{\Delta M}|$ carries essential information to derive the selection rules for dimer transitions:

$$\Delta S = S - S' = 0, \pm 1 \; ; \; \Delta M = M - M' = 0, \pm 1 \; . \tag{6}$$

The Q-dependence of the intensities of the observed multimer transitions displayed in Fig. 4 allows a proper peak identification. The intensity of the line $M_1$ follows the form-factor behavior for monomers described by Eq. (2), whereas the intensities of the lines $D_1$ and $D_2$ are governed by Eq. (5) for dimer transitions. We analyzed the dimer lines $D_1$ and $D_2$ in Fig. 2(b) on the basis of Eqs. (4-6) for both $Mn^{2+}$ and $Mn^{3+}$ ions. In order to cover all possible values of the ratio D/J, we introduce a parametrization scheme by putting D=Wy and J=W(1-|y|), where W is an energy-scale factor and -1≤y≤1. It follows that D/J=0 for y=0, while D/J=±∞ for y=±1. Out of a complete search for -1≤y≤1 (with steps Δy=0.1), agreement between the observed and calculated data could only be obtained for the parameter sets

$Mn^{2+}$: D=0.024(3) meV, J=-0.140(7) meV, (7)

$Mn^{3+}$: D=0.108(5) meV, J=-0.139(10) meV, (8)

*i.e.*, the Mn dimers are antiferromagnetically coupled. Similar values J=-0.16(3) meV [19] and J=-0.136 meV [20] were derived from magnetization and magnetic susceptibility data, respectively.

From the analysis of the dimer transitions both $Mn^{2+}$ and $Mn^{3+}$ are possible, thus we have to consider also the monomer transitions. The allowed monomer transitions are indicated in Fig. 3 by the arrows $M_1$ and $M_2$. Since the thermal populations of the first-excited monomer states ($|\pm 3/2\rangle$ for $Mn^{2+}$ and $|\pm 1\rangle$ for $Mn^{3+}$) are practically equal at T=1.7 K and T=6.0 K, the transition $M_2$ cannot be observed in the difference spectrum, so that only the monomer transition $M_1$ is accessible. According to Fig. 3, the monomer transition $M_1$ shown in Fig. 2(b) clearly has to be associated with $Mn^{2+}$. The resulting single-ion anisotropy parameter calculated from Eq. (1) is D=0.029(3) meV, in reasonable agreement with the value derived from the dimer transitions, see Eq. (7).

INS experiments were also performed for $Mn_xGa_{1-x}N$ with x=0.02 as shown in Fig. 2(c). The resulting energy spectra turned out to be very similar to those for x=0.04 displayed in Fig. 2(b), but with smaller intensity due to the lower Mn content. In the least-squares fitting procedure the linewidths of the three Gaussians were kept fixed at the values obtained for x=0.04. Obviously both the single-ion anisotropy parameter D and the exchange interaction J remain unaffected by the degree of Mn doping within the present experimental uncertainties.

The analyses of all our experiments performed for polycrystalline samples of the DMS $Mn_xGa_{1-x}N$ (x≤0.04) are consistent with a Mn valence of 2+. The $Mn^{2+}$ ions experience a substantial single-ion axial anisotropy which to our knowledge was neither addressed nor determined in all the studies carried out so far. The magnetic coupling between the $Mn^{2+}$ ions, resulting from N-bridged superexchange interactions, is antiferromagnetic and essentially independent of the Mn content x. This means that the injected holes are largely localized, so that the concentration of itinerant charge carriers is too low to generate a sizeable ferromagnetic component to the exchange coupling through a hole-mediated mechanism such as Zener's kinetic exchange interaction [21].

In fact, the concentration of mobile holes was measured to be $<10^{18}$ cm$^{-3}$ for crystalline $Mn_xGa_{1-x}N$ (x<0.1) [19]. A similar number was reported for the DMS compound $Mn_xZn_{1-x}Te$ (x≤0.05) where the exchange coupling determined by INS experiments experienced a marginal shift of not more than 1% due to the hole-mediated interaction [22].

The analysis of the neutron diffraction data did not provide a definite answer concerning the location of the dopant $Mn^{2+}$ ions in $Mn_xGa_{1-x}N$, although model II (see Table I) is favored for reasons given above. In order to arrive at a conclusive solution of the structure, investigations on single crystals or further neutron diffraction experiments extended to a larger Q-range (by using neutrons with wavelength of typically 0.5 Å) are highly desirable, which are expected to provide an improved discrimination of the models in terms of the $\chi^2$ test.

In conclusion, the realization of ferromagnetic DMSs with high Curie temperature relies on both the large moment of the substituted magnetic ions and carrier-induced ferromagnetic exchange interactions. For the crystalline compound $Mn_xGa_{1-x}N$ we have indeed a large magnetic moment (s=5/2), but our study showed that the nature of the exchange between the $Mn^{2+}$ ions is antiferromagnetic up to x=0.04. $Mn_xGa_{1-x}N$ samples with larger Mn content x have been obtained for thin films produced by various techniques such as molecular-beam epitaxy (MBE), metal organic chemical vapour deposition (MOCVD), and ion-assisted deposition (IAD). The latter method was used by Granville *et al.* [23] to prepare precipitate-free samples of $Mn_xGa_{1-x}N$ with Mn contents up to x=0.36. However, magnetic susceptibility measurements demonstrated that the exchange interaction between the $Mn^{2+}$ ions remains antiferromagnetic. In view of all these facts we conclude that the ferromagnetism reported in the literature for $Mn_xGa_{1-x}N$ (see references in Nelson *et al.* [4]) is likely due to the presence of either Mn clusters or ferromagnetic impurity phases. Nevertheless, alternative routes have been proposed to realize ferromagnetic DMSs which are based either on codoping with p-type elements such as Mg [24] or replacing the $Ga^{3+}$ ions by $Li^{1+}$ and $Zn^{2+}$ ions, so that the substitution of $Zn^{2+}$ by $Mn^{2+}$ ions is decoupled from carrier doping. The carrier concentration can then be controlled independently of Mn doping by adjusting the Li-(Mn,Zn) stoichiometry, as demonstrated for $Li_{1+y}(Mn_xZn_{1-x})As$ (0.05≤y≤0.2, 0.02≤x≤0.15) with Curie temperatures up to 50 K [25].

TABLE I. Neutron diffraction results obtained at room temperature for $Mn_{0.04}Ga_{0.96}N$ in the structure model $P6_3mc$ (No. 186) with lattice parameters a, b, and c. Ga is in the (2b)-position (2/3,1/3,0), and N and Mn in the (2b)-position (2/3,1/3,z). B denotes the isotropic displacement factor and p the occupation number. The reliability factors $R_n$ and $\chi^2$ are defined in Ref. [7]. The following constraints were applied: B(Ga)=B(Mn); p(Ga)+p(Mn)=1; p(N)=1. z(Mn) was kept fixed for model III at the value obtained from the Fourier analysis. For all models the impurity phase $MnN_y$ was refined in the space group I4/mmm with y=0.83 and resulting lattice parameters a=b=2.9762(5) Å and c=4.1311(13) Å.

|  | Model I | Model II | Model III |
|---|---|---|---|
| a (=b) [Å] | 3.19006(3) | 3.19011(3) | 3.19007(3) |
| c [Å] | 5.18608(5) | 5.18614(5) | 5.18608(5) |
| z(N) | 0.37790(10) | 0.37716(13) | 0.37782(10) |
| z(Mn) | 0(0) | 0.0597(44) | 0.18(0) |
| B(N) [Å$^2$] | 0.327(16) | 0.460(16) | 0.352(10) |
| B(Ga) [Å$^2$] | 0.321(23) | 0.175(19) | 0.281(13) |
| B(Mn) [Å$^2$] | 0.321(23) | 0.175(19) | 0.281(13) |
| p(Ga) | 1.012(8) | 0.945(4) | 0.981(2) |
| p(Mn) | -0.012(8) | 0.055(4) | 0.019(2) |
| $R_p$ | 3.15 | 3.10 | 3.13 |
| $R_{wp}$ | 3.96 | 3.88 | 3.93 |
| $R_{exp}$ | 2.54 | 2.54 | 2.54 |
| $\chi^2$ | 2.43 | 2.32 | 2.39 |

**FIGURE CAPTIONS**

FIG. 1. (Color online) Room temperature XANES spectra measured around the Mn K edge ($\approx$6.555 keV) for $Mn_xGa_{1-x}N$ (x$\approx$0.02 and x$\approx$0.04) as well as for the reference samples MnO ($Mn^{2+}$) and $Mn_2O_3$ ($Mn^{3+}$).

FIG. 2. (Color online) (a) Energy spectra of neutrons scattered from $Mn_xGa_{1-x}N$ (x=0.04) in the neutron energy-gain configuration at T=1.7 K and T=6.0 K. The error bars have the size of the symbols. The incoming neutron energy was 1.55 meV. (b) Difference energy spectrum I(T=6.0K)-I(T=1.7K) for $Mn_xGa_{1-x}N$ (x=0.04). (c) Difference energy spectrum I(T=6.0K)-I(T=1.7K) for $Mn_xGa_{1-x}N$ (x=0.02). The lines correspond to Gaussian least-squares fits as described in the text. The arrows mark the observed transitions.

FIG. 3. (Color online) Energy level splittings of magnetic monomers and dimers in $Mn_xGa_{1-x}N$ for $Mn^{2+}$ and $Mn^{3+}$. The energies are calculated from the parameters given in Eqs. (7) and (8). The full arrows mark the observed transitions displayed in Figs. 2(b) and 2(c). The dashed arrows refer to the remaining allowed transitions not observed in the INS experiments (the transition matrix element for $D_3$ is an order of magnitude smaller than for $D_1$ and $D_2$).

FIG. 4. (color online) Q-dependence of the neutron cross-section for $Mn^{2+}$ monomers and dimers. The lines denote the calculated intensities which are governed by Eq. (2) for monomer transitions and by Eq. (5) for dimer transitions with a Mn-Mn bond distance R=3.19 Å. The circles, squares, and triangles correspond to the intensities of the transitions $M_1$, $D_1$, and $D_2$, respectively, observed for $Mn_xGa_{1-x}N$ with x=0.04.

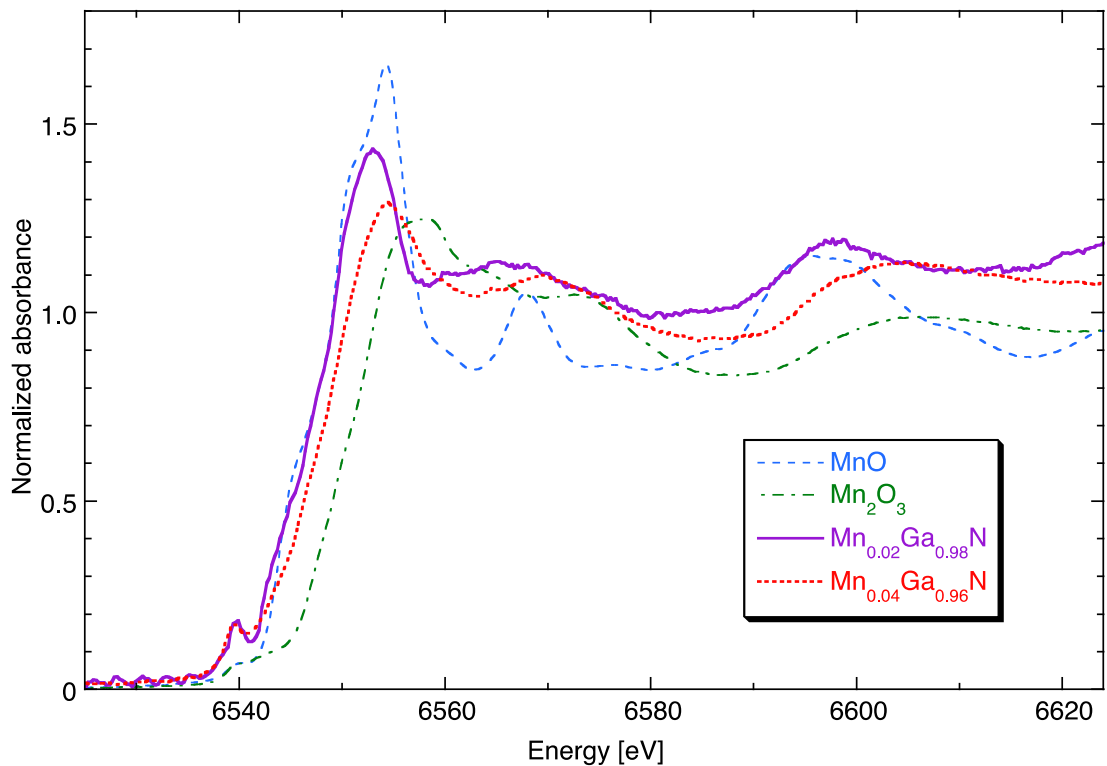

FIG. 1. (Color online) Room temperature XANES spectra measured around the Mn K edge (≈6.555 keV) for $Mn_xGa_{1-x}N$ (x≈0.02 and x≈0.04) as well as for the reference samples MnO ($Mn^{2+}$) and $Mn_2O_3$ ($Mn^{3+}$).

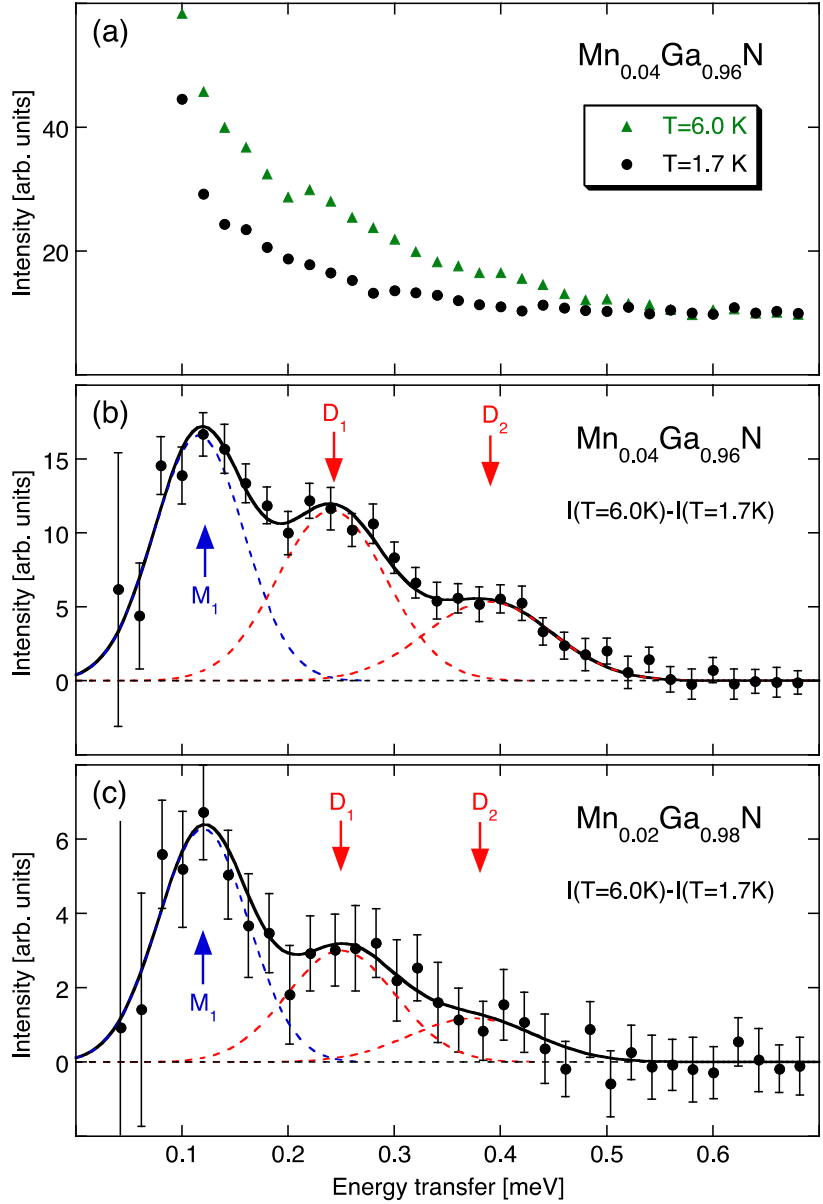

FIG. 2. (Color online) (a) Energy spectra of neutrons scattered from $Mn_xGa_{1-x}N$ ($x\approx0.04$) in the neutron energy-gain configuration at T=1.7 K and T=6.0 K. The error bars have the size of the symbols. The incoming neutron energy was 1.55 meV. (b) Difference energy spectrum I(T=6.0K)-I(T=1.7K) for $Mn_xGa_{1-x}N$ (x=0.04). (c) Difference energy spectrum I(T=6.0K)-I(T=1.7K) for $Mn_xGa_{1-x}N$ (x=0.02). The lines correspond to Gaussian least-squares fits as described in the text. The arrows mark the observed transitions.

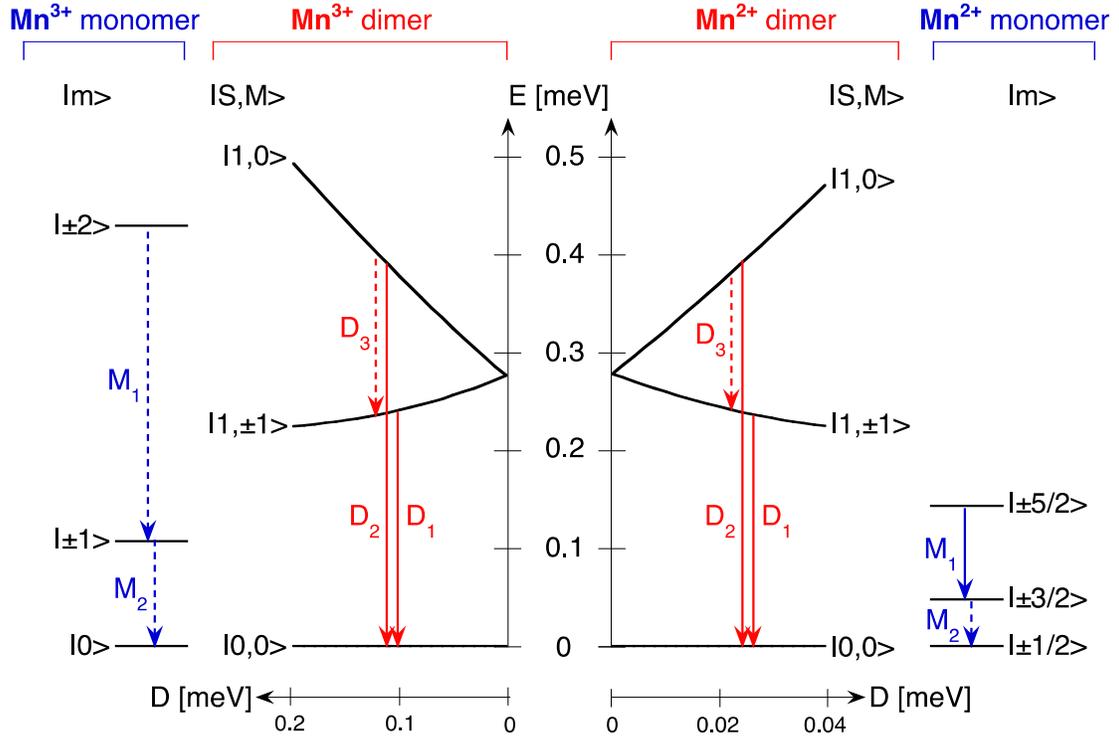

FIG. 3. (Color online) Energy level splittings of magnetic monomers and dimers in $Mn_xGa_{1-x}N$ for $Mn^{2+}$ and $Mn^{3+}$. The energies are calculated from the parameters given in Eqs. (7) and (8). The full arrows mark the observed transitions displayed in Figs. 2(b) and 2(c). The dashed arrows refer to the remaining allowed transitions not observed in the INS experiments (the transition matrix element for $D_3$ is an order of magnitude smaller than for $D_1$ and $D_2$).

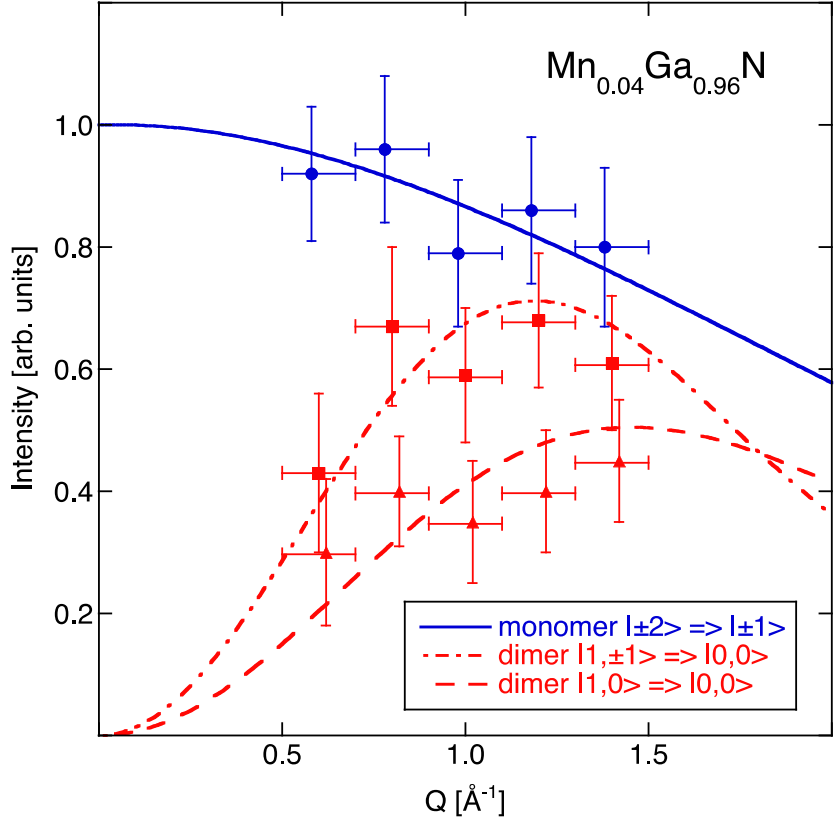

FIG. 4. (color online) Q-dependence of the neutron cross-section for $Mn^{2+}$ monomers and dimers. The lines denote the calculated intensities which are governed by Eq. (2) for monomer transitions and by Eq. (5) for dimer transitions with a Mn-Mn bond distance R=3.19 Å. The circles, squares, and triangles correspond to the intensities of the transitions $M_1$, $D_1$, and $D_2$, respectively, observed for $Mn_xGa_{1-x}N$ with x=0.04.